\documentclass{emulateapj}

\shorttitle{XMM Newton observation of Abell 1689}

\shortauthors{Andersson \& Madejski}

\begin{document} 

\submitted{Accepted for publication in \apj}

\title{Complex Structure of Galaxy Cluster Abell 1689:\\
Evidence for a Merger from X-ray data?}

\author{K.E. Andersson\altaffilmark{1,2} and G.M. Madejski\altaffilmark{1}}
\altaffiltext{1} {Stanford Linear Accelerator Center, 2575 Sand Hill Road,
Menlo Park, CA 94025, USA}
\altaffiltext{2} {Royal Institute of Technology (KTH), AlbaNova University Center (SCFAB), S-10691, Stockholm, Sweden}

\email{kanderss@slac.stanford.edu}

\begin{abstract}
  Abell 1689 is a galaxy cluster at $z=0.183$ where previous 
measurements of its mass using various techniques gave discrepant 
results. We present a new detailed measurement of the mass with 
the data based on X-ray observations with the European Photon 
Imaging Camera aboard the XMM-Newton Observatory, determined by using 
an unparameterized deprojection technique.  Fitting the total mass 
profile to a Navarro-Frenk-White model yields halo concentration 
$c=7.2^{+1.6}_{-2.4}$ and $r_{200}=1.13 \pm 0.21~ h^{-1}~\mathrm{Mpc}$,
corresponding to a mass which is less than half of what is found 
from gravitational lensing.  Adding to the evidence of substructure 
from optical observations, X-ray analysis shows a highly asymmetric 
temperature profile and a non-uniform redshift distribution implying 
large scale relative motion of the gas. A lower than expected gas 
mass fraction $f_{gas}=0.072~\pm~0.008$ (for a flat $\Lambda$CDM 
cosmology) suggests a complex spatial and/or dynamical structure.
We also find no signs of any additional absorbing component 
previously reported on the basis of the Chandra data, confirming the XMM 
low energy response using data from ROSAT. 
\end{abstract}

\section{Introduction} 
Galaxy clusters are the largest known gravitationally bound systems 
in the Universe.  The detailed analysis of the mass distribution 
of clusters is thus important in the process of understanding the 
large scale structure, and the nature of dark matter.  
The three main methods of measuring galaxy cluster masses:  
virial masses from velocity dispersions of cluster galaxies, 
X-ray imaging and spectroscopy of the intra-cluster medium (ICM) emission, 
and the gravitational lensing of background galaxies, have been 
found in recent years to be in disagreement for some clusters.  
Generally, the X-ray estimates are in good agreement with 
gravitational lensing for clusters with a high concentration 
of central X-ray emission (the so-called ``cooling flow'' 
clusters) but seemingly in disagreement for other, less centrally 
peaked objects \citep{all98}. To obtain the estimate of the 
total mass (including that due to dark matter) 
of a galaxy cluster from its X-ray emission -- commonly assumed to be 
from optically thin, hot plasma that subtends the space between galaxies -- 
it is necessary to make the assumption of hydrostatic equilibrium.  
This is appropriate of course only for clusters that have had time to 
relax into equilibrium and have not experienced any recent merger events.  
Generally, it is assumed that clusters with circular isophotes meet this 
criterion.  

Cold dark matter (CDM) hierarchal clustering is the leading theory 
describing the formation of large scale structure quite well.  
In particular, the numerical simulations such as \citet{nav97I} (NFW) 
successfully reproduce the observed dark matter halo profiles, which appear 
be independent of the halo mass, initial power spectrum of fluctuations, 
and cosmological model. However, observations often disagree with the 
numerical models.  For instance, one disagreement regards the rotation 
curves of dwarf elliptical galaxies which appear to be the result of a 
constant-density core whereas numerical simulations predict cuspy 
dark matter halo profiles \citep{moo99a}.  In addition, 
observations show fewer Milky Way satellites than predicted by the models 
\citep{kau93,moo99a}. 
Clearly, to understand the nature of galaxy clusters and the dark matter 
they consist of, it is important to measure the matter distribution 
in clusters via all available means.  Fortunately, 
there are two superb X-ray observatories, Chandra and XMM-Newton, 
featuring excellent angular resolution and exceptional effective 
area coupled with good spectral resolution, and those are well suited 
for detailed analysis of the X-ray emitting gas of galaxy clusters

\begin{figure*}
\plotone{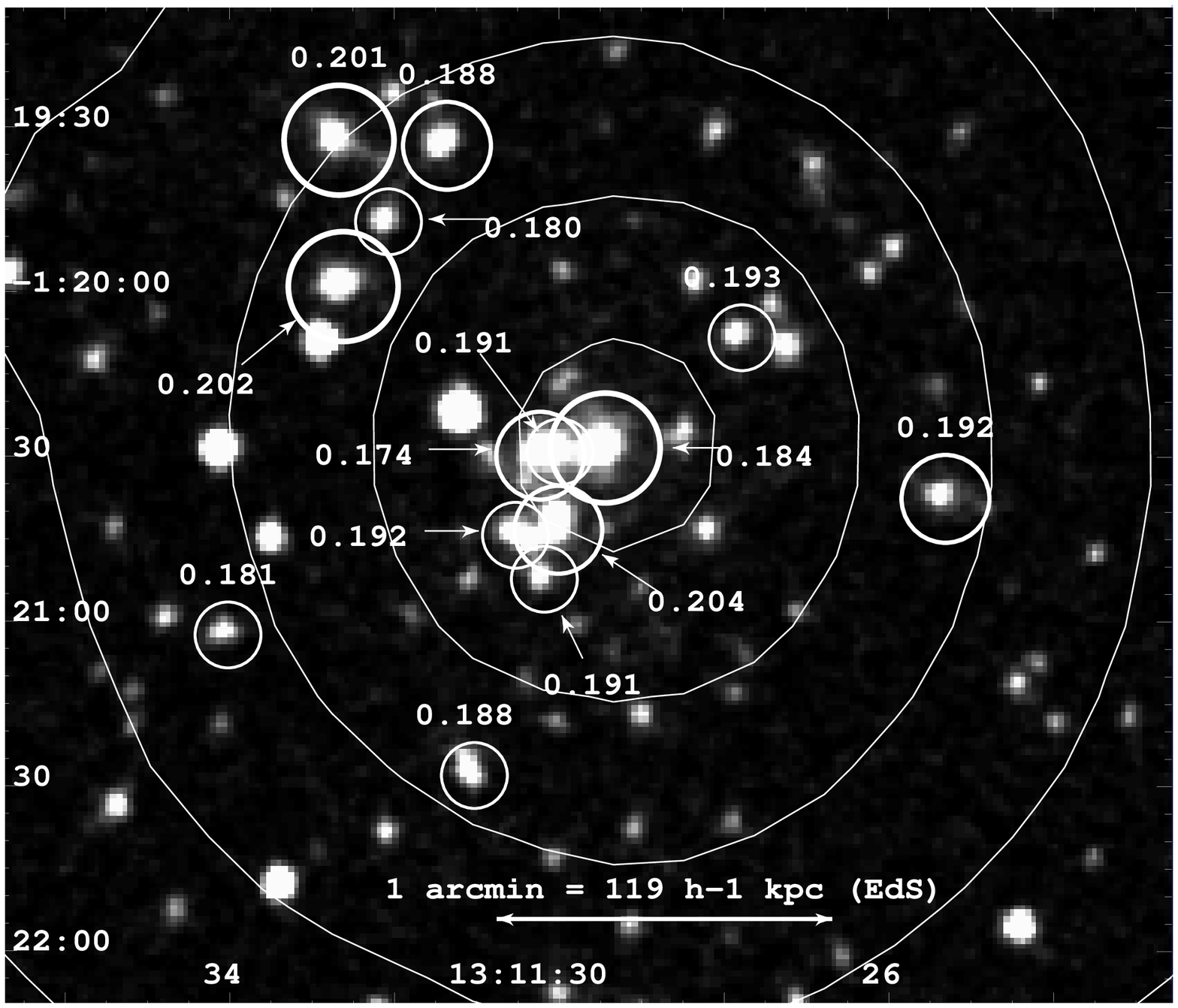}
\caption{\em R magnitude image of Abell 1689 from the STScI Digitized 
Sky Survey with all magnitude $R < 18$ cluster members within the 
central $2'$ region plotted over logarithmic X-ray contours from XMM 
{\sc Mos}. The bump toward northeast is due to a foreground point 
source. Circle sizes are proportional to galaxy R magnitudes. 
\label{figoptical}}
\end{figure*}

Abell 1689 is a cluster showing a large discrepancy among various 
mass determinations, and we chose it for a detailed study.  It 
is a rich cluster, $R=4$, without a pronounced cooling flow but 
with an approximately circular surface brightness distribution 
suggesting a relatively relaxed structure.  The large mass, 
$\sim 10^{15}~\mathrm{M_{\sun}}$ \citep{tys95}, and apparent symmetric 
distribution of Abell 1689 make it a suitable cluster for 
gravitational lensing measurements as well as X-ray measurements. 
However, the type of clusters that are believed to be the most relaxed 
have a cool central component of enhanced surface brightness. The
absence of such a component in Abell 1689 suggests that the cluster 
is not fully relaxed.  Also, the galaxy content of the cluster 
is unusually spiral rich for a cluster with high spherical symmetry 
and richness, with a galaxy type ratio E : S0 : Sp of approximately 
22 : 22 : 28 plus 25 unidentified galaxies \citep{duc02}.  
\citet{tea90I} measure a velocity dispersion of 
$\sigma_{1D}=2355^{+238}_{-183}~\mathrm{km~s^{-1}}$ for 66 cluster 
members, unusually high for a cluster of this temperature. Positions 
and redshifts from \citet{duc02} for all cluster 
members with red magnitude $R < 18$ and within $2 '$ of the brightest 
central galaxy are shown in Fig. \ref{figoptical} together with 
logarithmic X-ray intensity contours from XMM {\sc Mos} 1.

The gravitational lensing estimate from 6000 blue arcs and arclets calibrated 
by giant arcs at the Einstein radius of Abell 1689 gives a best fit 
power-law exponent of $n = -1.4 \pm 0.2$ for the projected density 
profile from $200~ h^{-1}~\mathrm{kpc}$ to $1 ~h^{-1}~\mathrm{Mpc}$ 
\citep{tys95}.  (Unless otherwise stated, we assume an 
Einstein-deSitter (EdS) cosmology with $\Omega_M = 1.0, 
\Omega_\Lambda = 0.0$ and $H_0 = 100 ~h~\mathrm{km~s^{-1}~ 
Mpc^{-1}}$.)  This is steeper than the profile of an isothermal 
sphere $(n = -1)$. The strong lensing analysis of two giant arcs 
directly gives $M_{2D}( < 0.10 ~h^{-1}~\mathrm{Mpc}) = 1.8 \pm 0.1 
\times 10^{14} ~h^{-1}~\mathrm{M_{\sun}}$ \citep{tys95}.

The mass profile derived from the deficit of lensed red galaxies 
behind the cluster due to magnification and deflection of background 
galaxies suggests a projected mass profile of $M_{2D}( < R) \approx 
3.5 \times 10^{15} (R~/~h^{-1}\mathrm{Mpc})^{1.3} ~h^{-1}~ 
\mathrm{M_{\sun}}$ for $R < 0.32~h^{-1}~\mathrm{Mpc}$, close to that 
of an isothermal sphere $(M_{2D} \propto R)$. The mass interior to 
$0.24 ~h^{-1}~\mathrm{Mpc}$ from this method is $M_{2D}( < 0.24 ~h^{-1}~ 
\mathrm{Mpc}) = 1.8 \pm 0.1 \times 10^{14} ~h^{-1}~\mathrm{M_{\sun}}$ 
\citep{tay98}.  Measurement of the distortion of the luminosity 
function due to gravitational lens magnification of background 
galaxies gives $M_{2D}( < 0.25 ~h^{-1}~\mathrm{Mpc}) = 0.48 \pm 0.16 \times 
10^{15} ~h^{-1}~\mathrm{M_\sun}$ \citep{dye01}.
Finally, the weak gravitational shear of galaxies in a ESO/MPG Wide 
Field Imager $33' \times 33'$ image gives a mass profile with best fit 
NFW profile with $r_{200} = 1.14 ~h^{-1}~\mathrm{Mpc}$ and $c = 4.7$ 
or a best fit SIS with $\sigma_{1D} = 1028^{+35}_{-42}~\mathrm{km~s^{-1}}$ 
\citep{clo01,kin02I}.

There is a good indication from the optical data that the cluster consists of 
substructures.  \citet{mir95} suggest a strong lensing model with two 
clumps in order to reproduce the positions of the brightest arcs.  
A larger mass clump $(\sigma_r = 1450~\mathrm{km~s^{-1}})$ is centered 
on the brightest cluster galaxy while a smaller clump 
$(\sigma_r = 700~\mathrm{km~s^{-1}})$ is located $1 '$ northeast of 
the main clump.  They arrive at a mass a factor 2 - 2.5 lower for their 
X-ray estimate compared to their gravitational lensing model.
\citet{gir97} identify two distinct substructures centered on 
redshifts $z=0.175$ and $z=0.184$ using positional and redshift data 
of cluster galaxies from \citet{tea90}, providing further evidence that 
the cluster is not relaxed. These clumps are also aligned 
in the southwest -- northeast direction but the locations do not agree 
with the ones of \citet{mir95}. Both structures are found to have 
$\sigma_r \sim 300-400~\mathrm{km~s^{-1}}$ yielding virial masses 
several times smaller than those derived from lensing and X-ray estimates.

Can the X-ray observations provide any evidence for a substructure in 
Abell 1689, and what are the implications on the mass inferred from the 
X-ray data?  In an attempt to answer this, we analyze XMM-Newton 
EPIC {\sc pn} and EPIC {\sc Mos} data to measure the mass profile 
of A1689 and to investigate the spatial structure of the cluster.  
\S 2 contains the details of the observations and data reduction with 
the XMM-Newton as well as with summary of the ROSAT, Asca and Chandra 
data;  \S 3 covers the methods of spectral fitting of the XMM data 
and presents the analysis of cluster asymmetry;  
\S 4 derives the mass and the slope of the mass distribution in the core;
and \S 5 presents the inferences about the structure of the cluster inferred 
from the spatial analysis.  The paper concludes with the 
summary in \S 6.

\section{Abell 1689 X-ray observations}
\subsection{XMM-Newton observation}
\subsubsection{Data preparation}
Abell 1689 was observed with XMM-Newton for 40 ks on 
December 24th 2001 during revolution 374. For imaging spectroscopy 
we use data from the European Photon Imaging Camera 
(EPIC) detectors {\sc Mos1}, {\sc Mos2} and 
{\sc pn}. Both {\sc Mos} cameras were operating in the Full Frame mode 
whereas {\sc pn} was using the Extended Full Frame mode. The Extended 
Full Frame mode for {\sc pn} is appropriate for studying diffuse sources 
since it has lower time resolution and so it is less sensitive to 
contamination from photons being detected during readout of the 
CCDs. These events (so called Out-Of-Time events) show up as streaks 
across the X-ray image and are especially bothersome when the goal of 
an observation is spatially resolved spectroscopy of diffuse 
sources. All cameras used the Thin filter during the observation.

EPIC background is comprised mainly of three components. The external 
particle background consists primarily of soft protons 
$(E_p < 100~\mathrm{keV})$ being funneled through the mirrors 
and causing a time variable flaring 
signal in the detector. The internal particle background is mainly 
due to high energy particles interacting with the detector material and 
causing a roughly flat spectrum with flourescent emission-lines 
characteristic to the detector material. This component 
varies over the detector surface. The third source of background is 
the cosmic X-ray background (CXB) which is roughly constant in 
time but varies over the sky.  

For all data reduction we use the software and calibration data 
implemented in XMM Science Analysis Software (SAS) 5.4.1. To exclude 
the events contaminated by proton flares, we produce light curves in 
the $10-15~$keV band where the true X-ray signal is low. We screen the 
data using a constraint on the total count rate of less than 1.5 ct 
s$^{-1}$ for {\sc Mos} and 1.1 ct s$^{-1}$ for {\sc pn} in this band, 
leaving an effective exposure time of 37 ks for {\sc Mos} and 29 ks for 
{\sc pn}. This screening corresponds to a limit on the count rate of 
approximately $2 \sigma$ above the quiescent period in the $0.3-10~$keV band.

\subsubsection{Vignetting correction}
The effective mirror collecting area of XMM-Newton is not constant 
across the field of view:  it decreases with increasing off-axis angle 
and this decrease is energy dependent.  This results in a position dependent 
decrease in the fraction of detected events and when doing imaging 
spectroscopy for extended sources, we need to account for this effect. 
By generating an Ancillary Response File (ARF) for each source spectrum 
region, using XMM SAS 5.4.1 command \verb1arfgen1 we calculate an average 
effective area for each region considered by us (see below).

\subsubsection{Background subtraction}
In order to correctly account for particle-induced and Cosmic 
X-ray Background, it would be optimal to extract a background 
spectrum from the same detector region collected at the same time as 
the source spectrum.  This is of course impossible, and the background 
can be taken from a source-free region of the detector 
(other than the target, but near it), or can be estimated using blank 
sky data.  We adopt the former method, noting that the background is 
not entirely constant over the field of view;  however, it can be 
assumed to be approximately constant with the exception of the 
fluorescent Cu line at $8~\mathrm{keV}$ in {\sc pn} which is the 
strongest contaminant emission line. Using this assumption, we can 
effectively subtract the particle background by extracting a spectrum 
in a source free region in the same exposure. For {\sc pn} data 
in the $7.8 - 8.2~\mathrm{keV}$ range is excluded due to the strong 
spatial dependence of the Cu internal emission.

Using an ARF generated for a source region on a background 
subtracted spectrum will not take into account that the spectrum used 
as background was extracted from a different region where the vignetting 
was higher, due to the larger off-axis distance of the background region. 
This will have the effect 
that the CXB component in our source spectrum will be under-subtracted 
and the net spectrum will contain some remnant of the CXB. 
The particle induced background however 
is not vignetted and therefore should leave no remnant.

To estimate the CXB component in our exposure we take the events outside 
of the field-of-view to be the particle background. Another spectrum is then 
extracted from a large source free area, located 
away from the cluster. The particle 
background is normalized and removed from this spectrum. The resulting spectrum is 
fitted to a broken power law model to determine the CXB flux. The incorrect 
vignetting correction of the CXB is found to cause an at most 2 \% 
over-estimation of the flux which is the case in the outermost annulus in 
our analysis (see below). However since the vignetting is energy-dependent, 
the incorrect vignetting may cause a small ($\sim 0.3$ keV) 
shift in temperature for the outermost region.  The overall uncertainty 
in the background is estimated to be at most a few percent.

We compare our background subtraction method with the method of 
using XMM standard blank sky data compiled by \citet{lum02};  
we find that both methods give similar fit parameters.  However, given the 
difficulty of normalizing the CXB, the different high-energy leftovers 
from proton-flare subtraction and the different internal particle 
background which occurs when using background data from a different 
exposure, we chose to use in-field (rather than blank-sky) 
background.  We find this is more robust in keeping the overall 
shape of the source spectrum uncontaminated. In all our analysis we 
use a background region of a circular annulus with inner and outer radii 
of $6'$ and $8'$ respectively. In fact, recent work by \citet{lum03} 
suggests that the in-field background method is probably more accurate. 
The clusters in the above paper, however, have smaller apparent angular 
size which makes the background method more reliable.

\begin{figure*}
\plottwo{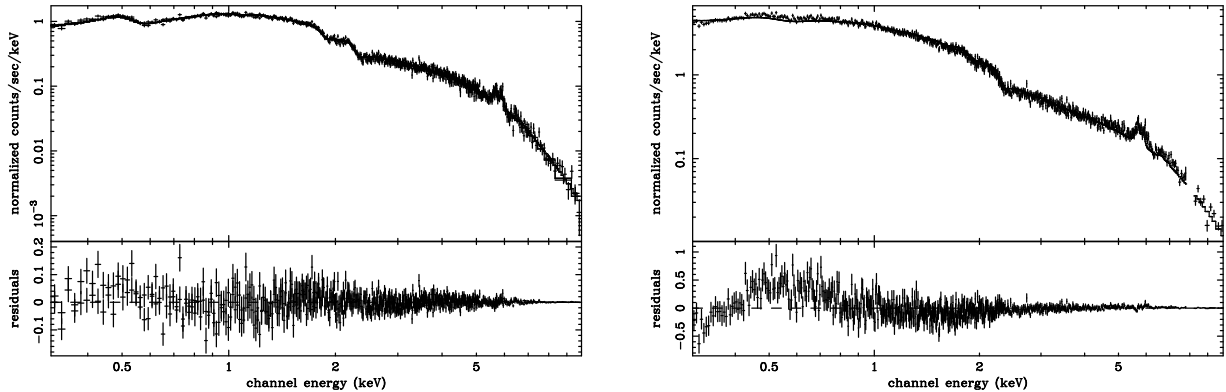}{f2b.eps}
\caption{\em Best fit MEKAL models for the XMM {\sc Mos} (Left) and 
{\sc pn} (Right) data from the central $3'$ region of Abell 1689. 
The absorbing column was fixed at the galactic value \citep{dic90}. 
\label{mospn}}
\end{figure*}

\subsubsection{Spectral fitting \label{fitting}}
In analyzing the vignetting corrected, background subtracted radial 
count-rate profile, we fit it to a conventional beta model, 
$S(r) \propto (1+r/r_c)^{-3 \beta + 0.5}$, where $S(r)$ is the source 
surface brightness at radius $r$. The fit 
gives $r_c = 91.2 \pm 0.7 ~h^{-1}~\mathrm{kpc}$ and $\beta = 0.72 \pm 0.01$ 
with $\chi^2/dof. = 46$ using data out to $700 ~h^{-1}~\mathrm{kpc}$. 
Clearly this model is not a very good fit;  we show it here only for 
completeness and comparison with previous work, and note that it is 
not used in the subsequent analysis. The bad fit above results from the 
fact that the cluster emission is more peaked 
in the core than the best fit beta model.  
To obtain a general idea of the properties of the 
cluster and compare this with previous results, we extracted 
spectra for the central $3 '$ $(356 ~h^{-1}~\mathrm{kpc})$ region, 
centered on the X-ray centroid at $13^h 11^m 29^s.4$  $-01^\circ 20' 28''$. 
This radius corresponds to $0.32~r_{200}$ or $3.9~r_{c}$. Background 
spectra were extracted from source-free regions from the same exposure. 
For XMM {\sc pn} we use single and double pixel events only 
whereas for XMM {\sc Mos} we also use triple and quadruple pixel events.

For spectral fitting we use the XSPEC \citep{arn96} software package.  
We fit the data in the $0.3-10.0~\mathrm{keV}$ range using the 
MEKAL \citep{mew85,mew86,kaa92,lie95} model for the optically thin 
plasma and galactic absorption. With the absorption fixed at the 
Galactic value, $N_H = 1.8 \times 10^{20}$ cm$^{-2}$ \citep{dic90}, 
and assuming a redshift 
$z = 0.183$ we arrive at a temperature of $9.35 \pm 0.17~\mathrm{keV}$ 
and a metal abundance of $0.27 \pm 0.03$ Solar for both {\sc Mos} 
cameras with $\chi^2 / \mathrm{dof.} = 951/824$. From the {\sc pn} 
camera we get $8.25 \pm 0.15~\mathrm{keV}$ and $0.23 \pm 0.03$ 
Solar with $\chi^2 / \mathrm{dof.} = 1424/863$. The best fit models 
with residuals can be seen in Fig. \ref{mospn} for {\sc Mos} (Left) and 
{\sc pn} (Right).
The {\sc pn} temperature is in 
disagreement with {\sc Mos} data and the reason for this effect can be 
seen from the residuals below 1 keV for {\sc pn} (Fig. \ref{mospn} (Right)). 
Fitting the {\sc pn}  data above 1 keV we get $9.33 \pm 0.20~\mathrm{keV}$ 
and $0.24 \pm 0.03$ Solar with $\chi^2 / \mathrm{dof.} = 892/722$, 
in better agreement with {\sc Mos}. The temperature from {\sc Mos} 
is in agreement with that found by Asca, $9.02^{+0.40}_{-0.30}~ 
\mathrm{keV}$ \citep{mus97}. This observation does not suffer from pile-up, 
nor is the low energy spectrum sensitive to background subtraction. 
The background uncertainties below 1 keV for this {\sc pn} spectrum are 
less than 1 \% whereas the {\sc pn} soft excess is sometimes 
higher than 10 \%. The excess is certainly not background related.  
It is possible that the {\sc pn} low-energy discrepancy can 
be due to incorrect treatment of charge collection at lower energies 
(S. Snowden, priv. comm.).  To resolve the discrepancy between 
{\sc Mos} and {\sc pn} we decided to compare with Asca GIS/SIS, 
ROSAT PSPC and Chandra ACIS-I data. 

\begin{figure*}
\plottwo{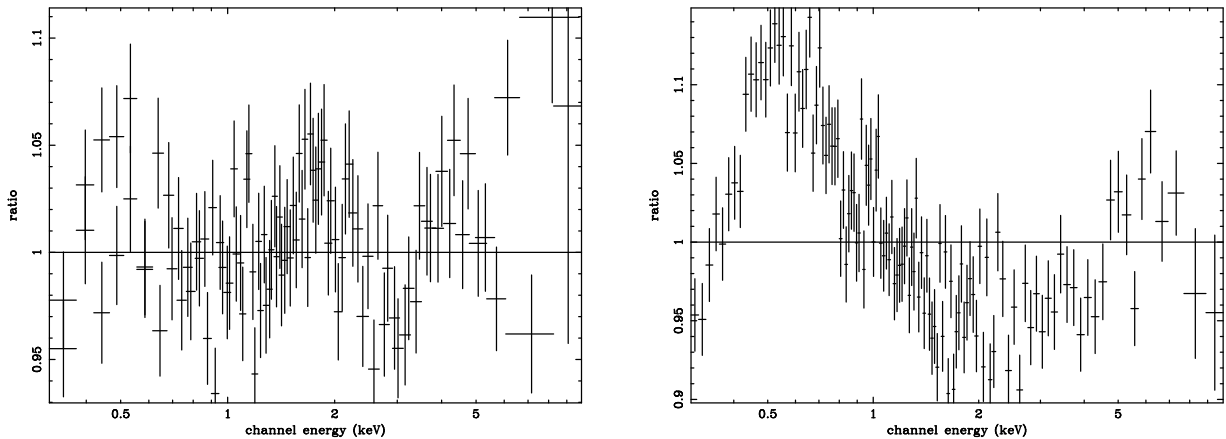}{f3b.eps}
\caption{\em The ratio of the XMM {\sc Mos} (Left) and {\sc pn} (Right) 
spectra of Abell 1689 to the best fit model determined from the ROSAT 
and Asca analysis. \label{MOSPNdisc}}
\end{figure*}

\subsection{ROSAT, Asca and Chandra observations \label{rosasc}}
Besides the discrepancy regarding the softest X-ray band for the 
Abell 1689 data between the {\sc Mos} and {\sc pn} detectors aboard 
XMM-Newton, the spectral fits to the Chandra data presented by 
\citet{xue02} imply a higher absorbing column, $6.7 \pm 1.5 \times 
10^{20}$ cm$^{-2}$ than the Galactic value of $1.8 \times 
10^{20}$ cm$^{-2}$ \citep{dic90}.  Since such excess absorption is 
not commonly detected in X-ray data for clusters, this requires further 
investigation.  To determine if there is indeed any additional component 
of absorption beyond that attributable to the Galactic column -- and 
assess the reliability of the softest energy band of the {\sc pn} vs. 
{\sc Mos} data -- we used the most sensitive soft ($< 1$ keV) X-ray data 
for this cluster obtained prior to the XMM-Newton observations, 
collected by the ROSAT PSPC.  The ROSAT observation conducted 
during July 18-24, 1992 (available from HEASARC) yielded 
13.5 ks of good data.  We extracted the ROSAT PSPC counts from a 
region $3'$ in radius, centered on the nominal center of X-ray 
emission.  For background, we selected a source-free region of the 
same PSPC image.  Using these data over the nominal energy range 
0.15 - 2.1 keV with the standard PSPC response matrix applicable to 
the observation epoch, we performed a spectral fit to a simple, 
single-temperature MEKAL model with soft X-ray absorption due to 
gas with Solar abundances, using the XSPEC package as above.  
In the fit we use metal abundances of 0.27 Solar obtained in the 
XMM {\sc Mos} fit above. While the limited bandpass of 
ROSAT PSPC precludes an accurate determination of the temperature 
(the best fit value is $kT = 4.3^{+1.2}_{-0.8}$ keV, 90\% confidence 
regions quoted), the PSPC data provide a good measure of the 
absorbing column:  the best value is $1.9 \pm 0.3 \times 10^{20}$ cm$^{-2}$, 
certainly consistent with the Galactic value.  We note that 
the difference between the temperature inferred from the PSPC fit 
and that obtained from the XMM-Newton data as above is a result 
of the limited bandpass of the PSPC, located much below the peak of the 
energy distribution of the cluster photons.  The measurement of 
the absorbing column, however, clearly indicates that the 
column inferred from the Chandra observation by \citet{xue02}
is not correct, and might be due to instrumental effects.  We
conclude that the absorbing column is consistent with the Galactic value. 

To obtain further constraints on the absorbing column, 
we also used the Asca GIS and SIS data together with the PSPC data 
for an independent constraint on the continuum radiation 
in the fitting process.  We performed standard extraction of data from 
all Asca detectors, also from a region from a region $3'$ in radius 
for the source, and a source-free region of the same image for 
background.  We performed a spectral fit simultaneously to data from 
the PSPC and four Asca detectors.  To account for possible flux calibration
differences, we let the normalization among all the different detectors 
run free. We used the energy range of $0.7-9.0~\mathrm{keV}$ and 
$0.5-8.5~\mathrm{keV}$ for the Asca GIS and SIS cameras respectively. 
Since Asca detectors (and in particular, the SISs) often return 
spectral fits with excess intrinsic absorption \citep{iwa99} 
we also let the absorbing column be fitted independently for the 
GIS, SIS and PSPC detectors. Temperature and metal abundances were tied 
together for all datasets in the fit. The optical redshift $z=0.183$ 
was used.  The joint fit of ROSAT and Asca data gives us the best 
fit temperature of $9.1 \pm 0.5$ keV and abundances of $0.25 \pm 0.06$ 
Solar, in agreement with the values quoted by \citet{mus97}, 
and the absorption (for the PSPC data) $1.75 \pm 0.08 \times 10^{20}$ 
cm$^{-2}$, in agreement with the Galactic HI 21 cm data. We shall use 
this value for the absorption in the subsequent analysis.

This best fit ROSAT/Asca model was compared with the data from the same 
region in the XMM-Newton cameras giving an unfitted reduced 
$\chi^2$ of 1.39 for {\sc pn} and 1.06 for both {\sc Mos} cameras 
combined. The ratio of these spectra to the ROSAT/Asca model can 
be seen in Fig. \ref{MOSPNdisc}. From this result 
we conclude that {\sc Mos} low energy response is more consistent 
with previous data and subsequently, we choose to ignore all {\sc pn} 
data below $1.1~\mathrm{keV}$. Re-fitting the XMM data from the above 
region using $0.3 - 10.0~ \mathrm{keV}$ for {\sc Mos} data and 
$1.1 - 10.0~ \mathrm{keV}$ for {\sc pn} leaving the absorbing column 
as a free parameter yields $n_H = 1.08 \pm 0.16 \times 10^{20}$ cm$^{-2}$, 
a temperature of $9.43^{+0.16}_{-0.15} ~\mathrm{keV}$ and metallicity of 
$0.26 \pm 0.02$ Solar. While this fitted value of absorption is formally 
inconsistent with the Galactic and ROSAT-inferred values, this is a relatively 
small difference, which might be due to the slightly imperfect calibration 
of the {\sc Mos} detectors or the assumption of isothermality made by us
for this fit (since $T$ and $n_H$ are correlated in the fitting procedure).  
We note that using the ROSAT value for absorption will 
give us a somewhat lower measure on the temperature (see below).  

Finally, we reduced the Chandra data for Abell 1689 using the most 
recent release of data reduction software;  specifically, due to the 
degradation of the Chandra ACIS-I low-energy response correction for 
the charge transfer inefficiency (CTI) is necessary, and we applied 
this to the Chandra data. As of Chandra data analysis software package 
CIAO ver. 2.3 this correction can be applied in the standard event 
processing. It is also necessary to account for the ACIS excess 
low-energy absorption due to hydrocarbon contamination. We used the 
\verb1acisabs1 code for the correction to the auxiliary response 
function. The event grades used in the ACIS analysis were GRADE=0,2,3,4 and 6.

\begin{figure*}
\plottwo{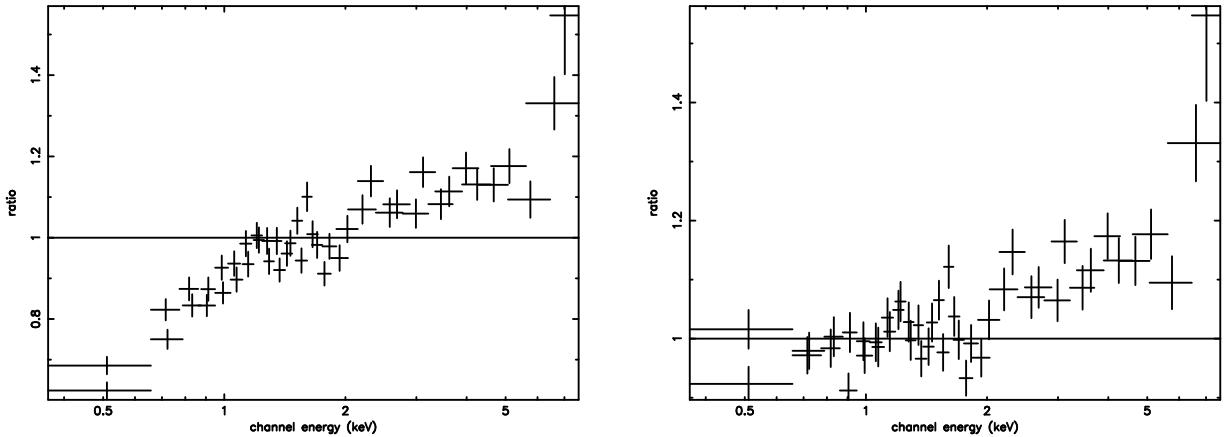}{f4b.eps}
\caption{\em The ratio of the Chandra {\sc ACIS-I} spectra of Abell 1689 
to the best fit model determined from the ROSAT and Asca analysis, 
before (Left) and after (Right) the acisabs correction. 
Data used for the spectrum include events with ACIS grades 0, 2, 3, 4, and 6.  
\label{chandra} }
\end{figure*}

The cluster was observed with the Chandra ACIS-I detector array at two 
separate occasions for $10~$ks each on 2000-04-15 and 2001-01-07. 
Spectra were extracted for the central $3 '$ region centered 
on the X-ray centroid at $13^h 11^m 29^s.4$  $-01^\circ 20' 28''$ 
following the CIAO 2.3 Science Threads for extended sources. 
Fig. \ref{chandra} shows the ratio of the combined Chandra data 
to the best fit model determined from ROSAT and Asca above. The 
left spectrum is before and the right after the \verb1acisabs1 
correction.  Fitting the corrected spectrum to an absorbed MEKAL 
model in the energy range 0.3 - 7.0 keV yields absorption of
$1.7 \pm 0.6 \times 10^{20}$ cm$^{-2}$, a temperature of 
$12.4 \pm 1.1$ keV and abundances of $0.34 \pm 0.10$ Solar using 
the optical redshift z=0.183. The discrepant absorption found by 
\citet{xue02} (using data 0.7 - 9.0 keV) is apparently corrected 
for by \verb1acisabs1 and the value is in agreement galactic absorption. 
However there is still a large discrepancy in temperature, 
which also can be seen from the high energy ends of Fig. 
\ref{chandra}. Part of this effect can possibly be attributed to 
the high energy particle background but more likely to 
uncorrected instrumental effects. Repeating all above steps 
for single pixel events only (GRADE=0) in order to achieve higher spectral 
accuracy gives us a best fit temperature of $7.2 \pm 0.4$ keV. 
We cannot account for the differences between the two Chandra data sets 
(using GRADE=0 vs. GRADE=0,2,3,4 and 6 events) nor between the results of 
the Chandra and XMM spectral fits.  We note here that the photon 
statistics resulting from 
the XMM observation is superior to that in the Chandra data, and since 
our analysis does not require the superior angular resolution of the 
Chandra mirror, we limit the analysis below to the XMM data.  

\section{Spectral analysis}
\subsection{Temperature and metallicity distribution for a spherically symmetric model}
To obtain a radial profile of cluster gas temperature, abundance, and 
density, we first make the assumption that the cluster is spherical and that 
above properties are only functions of radius.  For this, we divide the 
image of the cluster into 11 concentric annuli out to $5' 50''$ $(693 
~h^{-1}~\mathrm{kpc})$ centered on the X-ray centroid.  For each annulus, we 
extract spectra from all EPIC cameras, and we set the inner and outer 
radii of each region by requiring that each annulus contains at 
least 9000 counts per each {\sc Mos} camera and 13000 for {\sc pn}.  
This allows us to derive a reliable estimate of temperature in each 
region. Point sources with intensities greater than $3 \sigma$ over the 
average are excluded.  The outer radii of the annuli are as 
follows: $15''$, $25''$, $35''$, $47.5''$, $60''$, $75''$, $95''$, $125''$, 
$165''$, $230''$ and $350''$. 

Average cluster properties were determined using all annular spectra 
simultaneously out to $693 ~h^{-1}~\mathrm{kpc}$ 
($0.61~r_{200}$, $7.6~r_c$) using the 
same spectral fitting procedures as in Section \ref{fitting}.  
We use the optical redshift of $z = 0.183$ \citep{tea90} and the line of 
sight absorption of $N_H=1.75\times 10^{20}~\mathrm{cm^{-2}}$ as 
measured from ROSAT data, and also consistent with the Galactic 
column density (Section \ref{rosasc}). The mean temperature of the 
cluster is found to be $kT = 9.00^{+0.13}_{-0.12}~\mathrm{keV}$ and 
the mean abundance $0.25 \pm 0.02$ Solar. We note that leaving the 
redshift as a free parameter gives a best fit redshift $z=0.173 \pm 0.003$ 
(90\% confidence range) which is considerably less than measured 
via optical observations. Considering {\sc Mos} and {\sc pn} data 
separately gives $z=0.171 \pm 0.002$ and $z=0.178 \pm 0.003$ respectively.

In order to take into account the three-dimensional nature of the cluster we 
consider the spectrum from each of the 11 annuli to be a superposition 
of spectra from a number of concentric spherical shells intersected 
by the same annulus. The spherical shells have the same spherical radii 
as the projected radii of the annuli. We assume that each shell has a constant 
temperature, gas density and abundance. The volume for each annulus/shell 
intersection is calculated to determine how large a fraction of emission from each 
shell should be attributed to each annulus.  Assigning a spectral 
model to each spherical shell we can simultaneously fit the properties 
of all spherical shells.

In practice we use the method of \citet{ara02I} where we have a matrix 
of $11 \times 11$ MEKAL models with absorption in XSPEC. In the 
process of fitting the data in XSPEC, each datagroup (consisting of 
three data files:  {\sc pn}, {\sc Mos1} and {\sc Mos2} spectra for each 
annulus) is fitted using the same set of models.  For the 
central annulus (the one with zero inner radius so it's actually a circle), 
which will intersect 
all 11 spherical shells, we will need to apply 11 MEKAL models to 
represent these. This means that we have to apply 11 MEKAL models 
to each datagroup (annulus) with absorption where each model 
represents the properties of one spherical shell.  The normalization 
of each shell model is set to be the ratio of the volume that shell 
occupies in the cylinder that is the annulus/shell intersection 
to the volume it occupies in the central annulus. Abundance 
and temperature are tied together for the models representing the 
same spherical shell. Of course not all annuli intersect every 
shell, and for those shells not intersected by the annulus to which 
they are attributed, the model normalization will be zero. The matrix 
of MEKAL models is thus triangular and can be fitted directly to the 
spectra we have extracted from annular regions in the data.  
This approach allows for all data to be fitted simultaneously, 
and we do not have the problem with error propagation which occurs 
when starting to fit the outermost annulus and propagating inward 
subtracting contributions from each previous shell.

\begin{figure}
\plotone{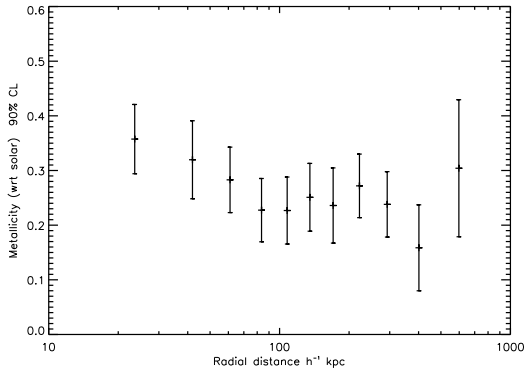}
\caption{\em The radial metallicity distribution of Abell 1689 as 
derived from the spectral deprojection fitting. \label{Abun}}
\end{figure}

\begin{figure}
\plotone{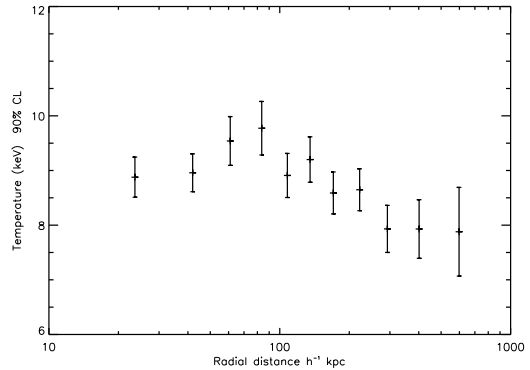}
\caption{\em The radial temperature distribution of Abell 1689 as derived 
from the spectral deprojection fitting. \label{Temps}}
\end{figure}

The metallicity profile (Fig. \ref{Abun}) from the deprojection shows 
signs of increasing abundance toward the center of the cluster.
In temperature (Fig. \ref{Temps}) we find an apparent decrease for 
large radii ($kT < 8~\mathrm{keV}$), an effect that has been seen in 
analysis of other clusters with XMM (see e.g. \citet{pra02}). Gas dynamic 
simulations of the formation of galaxy clusters also show a decline 
of temperature at large radii \citep{evr96}.
We do not find a significant cooling in the cluster core with the 
highest temperature ($kT \sim 9.5~\mathrm{keV}$) near the core radius:  
in fact, we will show in section \ref{xyana} that the temperature 
profile is not symmetric around the cluster center.  We note that for 
completeness, we also performed the above analysis with the best-fit value 
of the redshift inferred from the X-ray data alone, and while the 
exact values of temperature and elemental abundances are slightly 
different, about 0.2 keV lower for temperature, the general trends 
in the radial runs of the parameters are the same. 

The limited point spread function (PSF) of the XMM mirrors is a potential 
problem especially for the annuli located close to the center 
since those are not 
much larger than the PSF FWHM of $\sim 6''$. Some of the flux incident on 
the central (circular) region will be distributed over the outer annuli and 
vice versa. This flux redistribution will have the effect of smoothing 
out the measured temperature profile since all annuli will have some flux 
that originally belong in other annuli. This effect has been studied 
in detail by e.g. \citet{pra02} who find that correcting for the PSF 
redistribution gives a profile that is consistent with an 
uncorrected profile. Abell 1689 has a temperature profile without 
large temperature variations and no large central flux concentration. 
We conclude that the effect of flux redistribution in our case will 
be small and we do not attempt to correct for this. 
The PSF is also weakly energy dependent, and to quantify its possible 
effect on the observed temperature profile, we calculate 
the energy dependent flux loss from the central annulus and the 
effect on the central temperature. The difference in flux loss 
between various energy bands (ranging from 1.5 to 7.5 keV) 
for the on-axis PSF is $\sim 3 \%$. We find 
that for a cluster with an assumed temperature of 9 keV, 
this could give an error of the central temperature by 
at the most $0.5~\mathrm{keV}$. We note that this is a maximum 
difference since in practice, 
the flux gained from outer annuli could somewhat reduce this 
effect by working in the opposite manner.

The luminosity of the cluster in the $0.5 - 10.0~\mathrm{keV}$ 
band, calculated from the best fit model above (with $z = 0.183$) gives 
$L_{X~(EdS)} = 1.02 \times 10^{45} ~h^{-2}~\mathrm{ergs~s^{-1}}$ for 
an EdS ($\Omega_M=1.0$, $\Omega_\Lambda=0.0$) Universe or 
$L_{X~(CDM)} = 1.21 \times 10^{45} ~h^{-2}~\mathrm{erg~s^{-1}}$ for 
a CDM ($\Omega_M=0.3$, $\Omega_\Lambda=0.7$) Universe.  This corresponds 
to a bolometric luminosity of $L_{bol~(EdS)} =  1.63 \times 
10^{45} ~h^{-2}~\mathrm{erg~s^{-1}}$ or $L_{bol~(CDM)} =  1.94 \times 
10^{45} ~h^{-2}~\mathrm{erg~s^{-1}}$. All above values should include 10\% 
as the absolute calibration error of XMM. From Chandra analysis, \citet{xue02} 
find $L_{bol(EdS)}=1.66 \pm 0.64 \times 10^{45} ~h^{-2}~\mathrm{ergs~s^{-1}}$ 
whereas \citet{mus97} find  $L_{bol(EdS)}=1.77 \times 
10^{45} ~h^{-2}~\mathrm{ergs~s^{-1}}$ from Asca data, both in agreement 
with our results. This does not provide any new information regarding 
the location of Abell 1689 in the Luminosity-Temperature \citep{mus97} 
relation and it is still in a close agreement with the trend suggested by 
other clusters.

\begin{deluxetable}{llll}
\tabletypesize{\scriptsize}
\tablecaption{Comparison of the position of the 
cluster center of Abell 1689 from member galaxies, 
gravitational lensing and X-ray data \label{centers}}
\tablewidth{0pt}
\tablehead{
\colhead{Method} & 
\multicolumn{2}{c}{Center} &
\colhead{Ref.}\\
\colhead{} & 
\colhead{R.A. (J2000.)} &
\colhead{Dec. (J2000.)} &
\colhead{}
}
\startdata
X-ray (ROSAT) & $13^h 11^m 29^s.1$ & $-01^\circ 20' 40''$ & 1\\
X-ray (Chandra)  & $13^h 11^m 29^s.45$ & $-01^\circ 20' 28''.06$ & 2\\
Lensing & $13^h 11^m 29^s.6$ & $-01^\circ 20' 29''$ & 1\\
Optical & $13^h 11^m 29^s.44$ & $-01^\circ 20' 29''.4$ & 3\\
X-ray (XMM) & $13^h 11^m 29^s.4$ & $-01^\circ 20' 28''$ & 4\\
\enddata
\tablerefs{
(1) \citet{all98} ; (2) \citet{xue02}; (3) \citet{duc02};
(4) This study. 
}
\end{deluxetable}

\begin{figure*}
\plotone{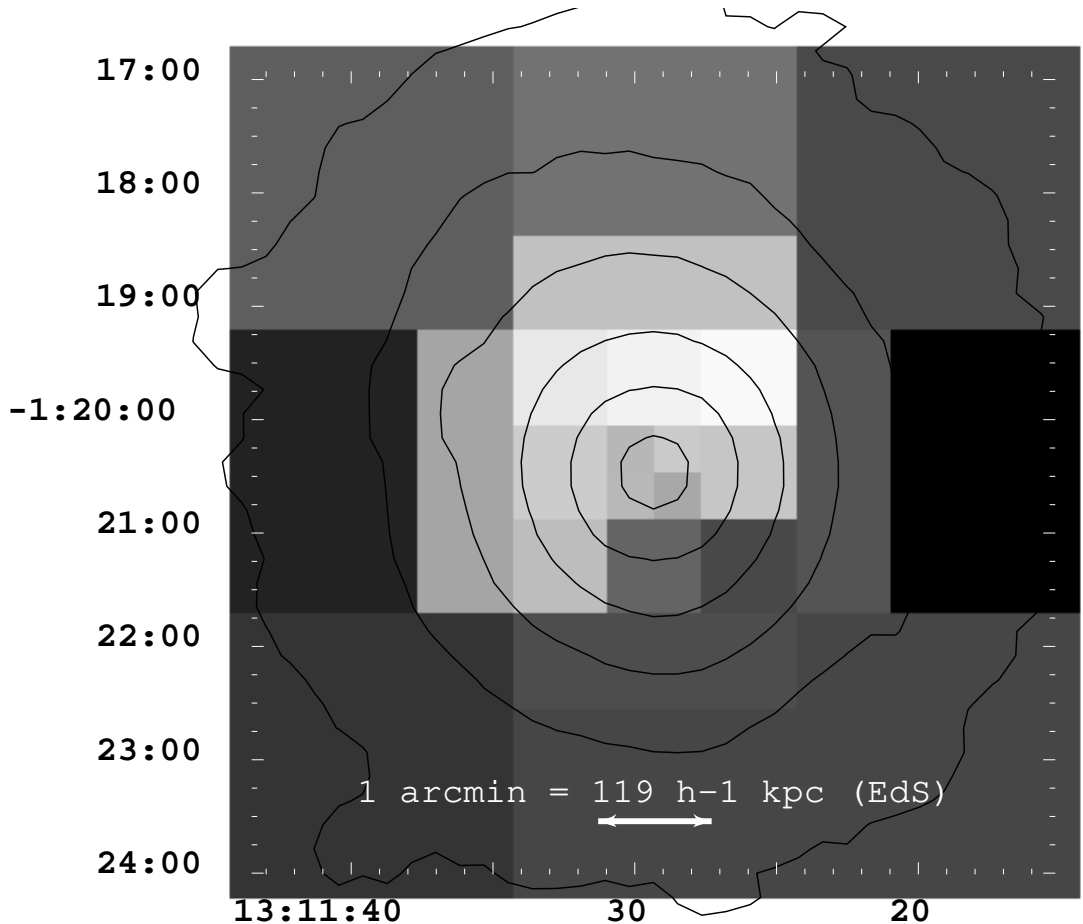}
\caption{\em Spatial distribution of spectral fit temperatures (6-10 keV) 
with superimposed logarithmic X-ray contours for A1689. 6 keV 
is marked as black, 10 keV is white, and intermediate temperatures
 are various shades of gray.  \label{Btemps}}
\end{figure*}

\subsection{Asymmetry analysis \label{xyana}}
With the good quality XMM data, it is possible 
to verify the result of \citet{xue02} 
that there is no discrepancy between the optical and X-ray centers 
of the cluster. We determine the center of X-ray emission for Abell 
1689 using XIMAGE command \verb1centroid1.
We also include a measurement of the lensing center 
from \citet{duc02} and the X-ray measurement from ROSAT by \citet{all98} 
(Table \ref{centers}). We find that all values agree within $3''$ except the 
ROSAT estimate, the offset of which we attribute to uncertainties in 
ROSAT HRI astrometry. This apparently perfect agreement among X-ray, 
lensing, and optical centers leads us to conclude that the ICM density 
peak and the central dominant galaxy is probably located at the bottom 
of the dark matter potential well.  Still, 
the apparent non-uniformity of the ICM radial temperature distribution 
as well as the offset of optical and X-ray redshifts prompted us to 
analyze the spatial structure of the cluster. In Fig. \ref{Btemps} 
we show the spatial temperature distribution. Spectra were extracted 
in rectangular regions and fitted using the same method as in 
Section \ref{fitting}. The temperature in the figure scales linearly from 
6 keV (black) to 10 keV (white). The errors on the temperature of the 
inner 16 regions are $\sim 0.5~\mathrm{keV}$, while the errors on the 
outer 8 regions are $\sim 1.0~\mathrm{keV}$. We clearly see an asymmetry 
in the temperature around the cluster center with an overall increase 
toward the northeast.

\begin{figure*}
\plotone{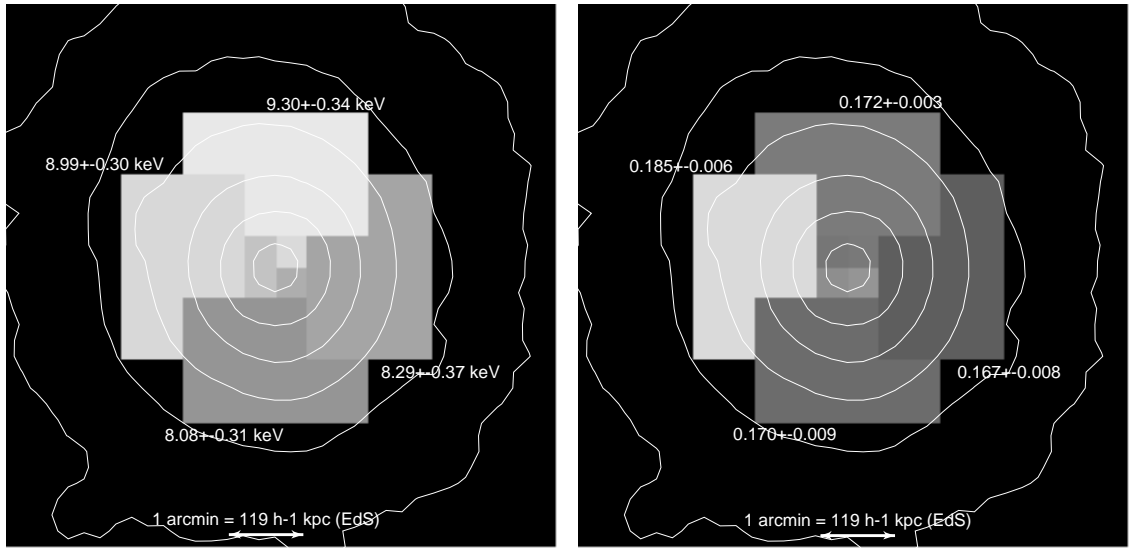}
\caption{\em The spatial distribution of temperature (left) and redshift 
(Right) in the central region of Abell 1689 with 90\% confidence intervals. 
\label{URDL}}
\end{figure*}

To check the consistency of these results and to increase our accuracy, 
we re-group the data in larger spatial regions and perform a fit using 
the same above procedure. We first fit the data keeping the redshift 
frozen at the optical value. Temperatures derived from this fit are 
shown in Fig. \ref{URDL} (Left).  Thereafter, we leave the redshift 
as a free parameter and re-fit the data. The fitted redshifts for 
these regions are shown in Fig. \ref{URDL} (Right). All errors 
in Fig. \ref{URDL} are 90\% confidence limits.   In the temperature 
map, we see a clear discrepancy between the northern and southern part 
of the cluster with a hint of a temperature gradient in the southwest -- 
northeast direction.  The redshift map reveals a high redshift structure 
to the east at $z=0.185 \pm 0.006$ separated from the rest of the 
cluster at $z \sim 0.17$. Analyzing the {\sc Mos} and {\sc pn} data 
separately for this high redshift region gives a broad 
minimum in  $\chi^2$ at $z=0.187 \pm 0.008$ for {\sc Mos} whereas {\sc pn} 
shows several minima in the $z=0.165 - 0.200$ range.

This region may indicate a subcluster falling inward away from the observer.  
This is further supported by the optical data, indicating that 
there are also high-redshift giant elliptical galaxies in this region 
(Fig. \ref{figoptical}):  it is interesting to speculate if this is 
actually the remains of a cluster core?  Especially intriguing is the 
apparent coincidence between the smaller subcluster as suggested by 
strong lensing \citep{mir95} and our high redshift gas region 
approximately $1'$ northeast of the main cluster.  
Another possibility is that the redshift variation is due to large 
bulk motions of the intra-cluster gas. It has been shown in gas dynamic 
simulations that clusters with apparently relaxed X-ray profiles can 
have complex gas-velocity fields and be far from relaxed \citep{evr96}.  
This kind of motion could give rise to non-thermal emission from shocks 
etc. In our analysis we cannot distinguish between the two possibilities 
of bulk motion and subclustering.  

This measurement of non-uniform redshift distribution of the 
X-ray emitting gas is important, and to verify if it could be due to 
instrumental effects, we investigated this in more detail.  
The data in the regions above are from different CCD chips in the 
{\sc pn} data whereas for {\sc Mos} all data are from the same chip. 
Since the {\sc pn} camera provides about half of the data, we want to 
verify that there are no gain shifts between the CCDs, which, if 
present, could easily cause such an effect.  Most of the cluster 
emission is on the CCD chips 4 and 7. To test for any possible gain offset, 
we extracted the data from each of the chips individually to verify 
the position of the internal fluorescent $\mathrm{Cu K}\alpha$ line, 
with the dominant component at $8.0478~\mathrm{keV}$.  
The spectra in the range $7.8 - 8.2~\mathrm{keV}$ are fitted to a 
Gaussian profile:  both datasets yield essentially the same peak 
position of $8.051 \pm 0.001~ \mathrm{keV}$. This corresponds to a 
possible artificial redshift offset of maximum $\Delta z = 0.0005$.  
To explain the difference in the offset measured by us as an 
instrumental effect, we would have to have an offset (say at the Fe K 
line at 6.7 keV) of 62 eV, and not on the order of 1 eV, as inferred 
from the Cu K instrumental line.  Hence we conclude that there is no 
gain shift that could alter our redshift measurements between {\sc pn} 
CCDs 4 and 7.  According to XMM calibration documentation \citep{kir03} 
the magnitude of calibration errors for {\sc pn} \& {\sc Mos} 
should be no larger than 10 eV.

\section{Mass Profile}

\subsection{Mass calculation}
If we assume that the cluster is spherical with a smooth static 
gravitational potential and that the intra-cluster medium is a 
pressure-supported plasma, we can employ the hydrostatic equilibrium 
equation. The circular X-ray isophotes (Fig. \ref{figoptical}) 
generally indicate that a cluster is in dynamical equilibrium. 
The hydrostatic equation can be written as \citep{sar88} :
\begin{equation}
M(r)=-\frac{k T_g(r)~r}{G m_p \mu} \left ( \frac{\mathrm{d}~ln~T_g(r)}{\mathrm{d}~ln~r}+ \frac{\mathrm{d}~ln~\rho_g(r)}{\mathrm{d}~ln~r} \right )
\label{mass}
\end{equation}
where $M(r)$ is the enclosed total gravitating mass enclosed within 
a sphere of a radius $r$, $T_g(r)$ and $\rho_g(r)$ are temperature 
and density of the ICM at $r$, $\mu$ is the mean particle weight 
and $m_p$ is the proton mass. Using the temperature and normalizations 
from the spectral deprojection fitting we calculate the total 
gravitating mass. Errors are treated by error propagation. 

The mass distribution is fitted to a singular isothermal sphere (SIS)
\begin{equation}
M(r)=\frac{2 \sigma_r^2 r}{G}
\label{massSIS}
\end{equation}
where $\sigma_r$ is the 1-dimensional velocity dispersion, which is used here 
for a comparison with previous mass estimates. 

The predicted density profile from CDM hierarchal clustering 
according to \citet{nav97} for dark matter halos is
\begin{equation}
\frac{\rho(r)}{\rho_{crit}(z)} = \frac{\delta_c}{(r/r_s)(1+r/r_s)^2}
\label{eqNFW1}
\end{equation}
where $\rho_{crit}(z)$ is the critical energy density at halo 
redshift $z$ and $\delta_c$ is characteristic density defined by
\begin{equation}
\delta_c = \frac{200}{3}\frac{c^3}{[ln(1+c)-c/(1+c)]}
\label{eqNFW2}
\end{equation}
where $c=r_{200}/r_s$ is the concentration of the halo defined as the 
ratio of the virial radius $r_{200}$ to $r_s$, which in turn is a 
characteristic radius in the NFW model. The critical density 
at redshift $z$ for a flat ($\Omega_0 = 1$) Universe is 
\begin{equation}
\rho_{crit}(z)=\frac{3H_0^2}{8\pi G} [\Omega_M (1+z)^3+\Omega_\Lambda ]
\label{rhocrit}
\end{equation}
where $H_0$ is the Hubble constant, $\Omega_M$ and $\Omega_\Lambda$ are 
the current contributions of matter and vacuum energy respectively to 
the energy density of the Universe.

More recent numerical studies suggest a steeper core slope and a sharper 
turn-over from small to large radii \citep{moo99}.  Both models can be 
generalized as
\begin{equation}
\rho(r) = \frac{\rho_0}{(r/r_s)^\gamma [1+(r/r_s)^\alpha]^{(\beta - \gamma)/\alpha}}
\label{eqGeneral}
\end{equation}
\citep{zha96}, where $\gamma$ and $\beta$ characterize the density 
slope at small and large radii respectively whereas $\alpha$ 
determines the sharpness in the turn-over. Most studies agree on 
$\beta = 3$ but the value of $\gamma$ is still being debated. The NFW 
and Moore profiles 
fit into the parameter space $(\alpha, \beta, \gamma)$ as $(1, 3, 1)$ 
and $(1.5, 3, 1.5)$ respectively.

\begin{figure}
\plotone{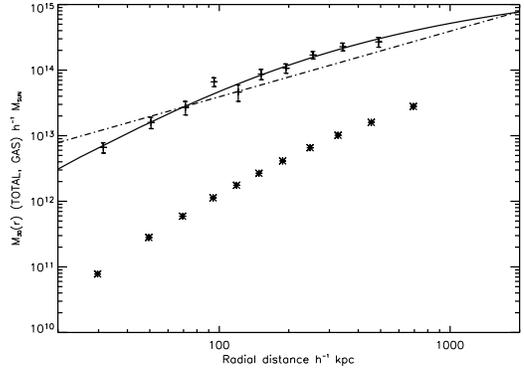}
\caption{\em Spherical mass profile of Abell 1689 (pluses) with 
best fit NFW model (solid line) and singular isothermal sphere model 
(dot-dashed). The singular isothermal sphere is clearly not a very 
good fit to the data (see also Table \ref{bestfit}). For comparison, 
the mass of the intra-cluster gas is included (asterisks). \label{Mass3D}}
\end{figure}

We choose to fit our data to the NFW model (Eq. \ref{eqNFW1}) which, 
when integrated over $r$, yields 
\begin{equation}
M(r)=M_0 \times [\ln(1+r/r_s) + (1+r/r_s)^{-1} -1]
\label{massNFW}
\end{equation}
where $M_0=4 \pi \rho_c \delta_c r_s^3$. We find that the data give 
the best fit for the NFW model with $c=7.2^{+1.6}_{-2.4}$ and 
$r_{200}=1.13 \pm 0.21 ~h^{-1}~\mathrm{Mpc}$ whereas the SIS 
fit gives $\sigma_r=918 \pm 27~\mathrm{km~s^{-1}}$.  
The total mass data and models are shown in Fig. \ref{Mass3D} 
together with the mass of the X-ray emitting gas $M_{gas}$.  
Model parameters are summarized in Table \ref{bestfit}.
For a cosmology with $\Omega_M=0.3$ and $\Omega_\Lambda=0.7$, 
the best fit NFW model changes to $c=7.7^{+1.7}_{-2.6}$ and 
$r_{200}=1.31 \pm 0.25 ~h^{-1}~\mathrm{Mpc}$. 

Another cosmologically important quantity for clusters is the fraction of the 
total mass that is in the X-ray emitting gas, 
$f_{gas}=M_{gas}/M_{TOT}$.  \citet{all03} analyze 
data from the Chandra Observatory for 
10 dynamically relaxed clusters between $z=0.09$ and $z=0.46$ and measure 
an average redshift independent $f_{gas} = 0.108 \pm 0.014$ at the 
$r_{2500}$ radius (where the total mass density is 2500 times the 
critical density at the redshift of the cluster).  The cosmology where 
this is valid is a flat $\Lambda\mathrm{CDM}$ cosmology with 
$\Omega_M=0.291^{+0.040}_{-0.036}$ assuming 
$\Omega_b ~h^2 = 0.0205 \pm 0.0018$ and $h=0.72 \pm 0.08$. 
The error on $f_{gas}$ above is the rms dispersion of the 
\citet{all03} sample which is comparable to the individual errors on 
$f_{gas}$.  With our best fit NFW model in the above cosmology we find 
$f_{gas}=0.072 \pm 0.008$ for Abell 1689 at the $r_{2500}$ radius. 
This is lower than all the clusters in the \citet{all03} sample and 
significantly lower than the mean. We find that in our estimate 
$f_{gas}$ has not converged to a constant and this might help explain 
part of the discrepancy. However, it may be the case that for many clusters 
$f_{gas}$ does not converge until well beyond $r_{2500}$.

\begin{deluxetable*}{cccc}
\tabletypesize{\scriptsize}
\tablecaption{Best fit models for the total mass profile of Abell 1689 \label{bestfit}}
\tablewidth{0pt}
\tablehead{
\colhead{Model} & 
\colhead{Range~$[h^{-1} \mathrm{kpc}]$} &
\colhead{Parameters} &
\colhead{$\chi^2$ (dof.)}
}
\startdata
$\mathrm{NFW}_{EdS}$\tablenotemark{a} & $20 - 500$ & $c=7.2~ r_{200}=1.13 ~h^{-1}~\mathrm{Mpc}$ &
7.64 (8) \\
$\mathrm{NFW}_{CDM}$\tablenotemark{b} & $20 - 500$ & $c=7.7~ r_{200}=1.31 ~h^{-1}~\mathrm{Mpc}$ & 
7.64 (8) \\
SIS & $20 - 500$ & $\sigma=918~\mathrm{km~s^{-1}}$ & 
61.4 (9) \\
POW & $20 - 90$ & $\gamma=1.73 \pm 0.34$ & 
\tablenotemark{*} \\
POW & $200 - 500$ & $\gamma=0.72 \pm 0.32$ & 
\tablenotemark{*} \\
\enddata

\tablenotetext{a}{EdS refers to a flat cosmology with $\Omega_M=1$ and $\Omega_\Lambda=0$}
\tablenotetext{b}{CDM refers to a flat cosmology with $\Omega_M=0.3$ and $\Omega_\Lambda=0.7$}
\tablenotetext{*}{These fits are with one degree of freedom only and hence we do not state $\chi^2$ for these.}
\tablecomments{The singular isothermal sphere (SIS) and powerlaw (POW) models are not quoted for different cosmologies since they are independent of cosmological parameters.}

\end{deluxetable*}

Comparing the mass and temperature at $r_{2500}$ of Abell 1689 to the 
M-T relation derived for a set of relaxed clusters \citep{all01} shows a 
low mass for the temperature of Abell 1689. The M-T relation predicts 
$H(z)/H_0 \times M_{2500} = 4.5 \times 10^{14} \mathrm{M_{\sun}} (\pm 10 \%)$ for 
a $9~\mathrm{keV}$ cluster where $H(z)$ is the Hubble constant at the 
redshift of the cluster. For Abell 1689 we find 
$H(z)/H_0 \times M_{2500} = 2.4 \times 10^{14} \mathrm{M_{\sun}} (\pm 15 \%)$, 
significantly lower. The above values were derived assuming a flat 
$\Lambda\mathrm{CDM}$ cosmology with $\Omega_M=0.3$, $\Omega_\Lambda=0.7$ 
and $h = 0.7$. The unusually low mass may be due to the fact that the 
mass of Abell 1689 seems to increase steadily beyond $r_{2500}$. However, we note that a lower mass (than would be predicted by the M-T relation) 
is not an uncommon feature for unrelaxed clusters (cf. \citet{smi03}).

For completeness, we note that the calculated total mass includes 
the ICM and galaxy mass contributions as well as the dark matter. 
The proper way of fitting the NFW model would be to subtract these 
contributions prior to performing the fit. The NFW model for the 
actual dark matter halo is not used in this paper;  we only calculate 
the total mass profile.

\subsection{Core slope}
The mass data were fitted to a simple power law ($M(<r) \propto r^{\gamma}$) 
in the ranges $20 - 90 ~h^{-1}~\mathrm{kpc}$ and $200 - 500~  h^{-1}~
\mathrm{kpc}$.  We find the best fit of the slope of the matter profile to 
to be $1.73\pm 0.34$ and $0.72\pm 0.32$ for small and large radii, 
respectively. This corresponds to total mass density slopes 
($\rho \propto r^{\alpha}$) of $\alpha=-1.27 \pm 0.34$ and 
$\alpha=-2.28 \pm 0.32$, in good agreement with what is expected from 
numerical simulations of CDM hierarchal clustering. We note here that 
\citet{bau02} have measured the density profile of Abell 1689 for 
$r < 100~  h^{-1}~\mathrm{kpc}$ using Chandra data and found 
$\alpha \sim -1.3$. 

We do not observe a flattening of the core density profile but find 
a slope close to the core to be somewhere in between the preferred 
Moore and NFW profiles. Thus we do not claim to be able to detect 
nor dismiss any kind of modification to standard cold dark matter. 
Nonetheless, the X-ray data imply an upper limit on the self-interaction 
of dark matter;  see, e.g., the discussion in \citet{ara02}.  
Such comparisons are more meaningful for cooling-flow clusters 
which are presumably more relaxed objects.

\section{Discussion}

\subsection{Comparison with Lensing}
\begin{figure}
\plotone{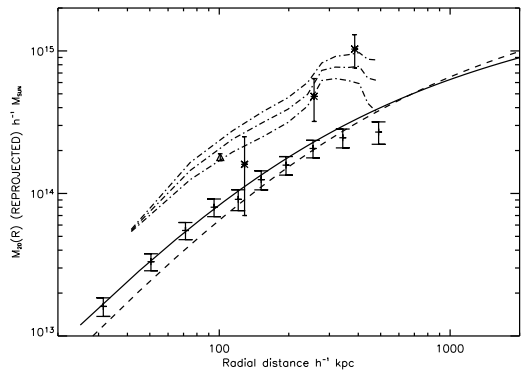}
\caption{\em Unparameterized projected mass distribution, assuming that 
our the outermost shell is the limit of the cluster mass (pluses with 
errorbars), plotted with reprojected best fit NFW model (solid line).  
Also shown are gravitational lensing results from strong lensing (triangle), 
distortion of background galaxy luminosity function (asterisks), deficit 
in number counts of red background galaxies (dot-dashed lines) and 
projected best fit NFW model from weak gravitational shear (dashed line).  
\label{Mass2D}}
\end{figure}

To compare our results with those obtained from gravitational lensing, 
we reprojected our derived mass distribution into a two dimensional 
distribution by summing up the contributions from each shell along 
the line-of-sight.  Of course this method assumes that the 
outermost shell is the absolute limit of the cluster, and we recognize 
that this will not be entirely accurate.   Therefore we also include 
our projected best-fit NFW model which has an analytic expression. 
Reprojected mass and NFW model are shown in Fig. \ref{Mass2D} where 
pluses with error bars are the reprojected mass, 
the solid line is our NFW model, the triangle is the strong lensing 
result \citep{tys95}, the asterisks are lensing magnification results 
measured by the distortion of the background galaxy luminosity function 
\citep{dye01}, the dot-dashed lines are lensing magnification results 
measured by the deficit of red background galaxies \citep{tay98}, and 
the dashed line is the best fit NFW model from weak gravitational 
shear analysis \citep{clo01,kin02}. The comparison of our results 
with measurements from gravitational lensing is summarized in 
Table \ref{oldres}.  

The X-ray mass appears to be in good agreement with that derived from 
weak gravitational shear but cannot be reconciled with the strong 
lensing data nor with the data from gravitational lens magnification.  
This discrepancy in lensing was noted earlier by \citet{clo01} as well as 
\citet{kin02} who cannot find an agreement using any realistic corrections. 
The high velocity dispersion measured by \citet{tea90} and the apparent 
grouping of galaxies along the line of sight \citep{gir97} may indicate 
that this cluster is not as regular as we expect from its smooth 
circular surface intensity. The two component model of \citet{mir95} 
also strongly suggests non-uniformity. This might explain 
the discrepancies between mass estimates from gravitational lensing and 
from X-rays. A possibility could be that this is a cluster undergoing 
a major merger, where a sub-clump close to or along the line-of-sight 
is being stripped from gas or just has very low X-ray luminosity.

\begin{deluxetable*}{cccc}
\tabletypesize{\scriptsize}
\tablecaption{Comparison of previous mass estimates from gravitational lensing to our X-ray estimate of Abell 1689 \label{oldres}}
\tablewidth{0pt}
\tablehead{
\colhead{$M_{2D}~[10^{15} ~h^{-1}~\mathrm{M_{\sun}} ]$} & 
\colhead{Radius~$[h^{-1}~\mathrm{Mpc}]$}   & 
\colhead{Ref.} &
\colhead{Type of Measurement}
}
\startdata
$0.18 \pm 0.01$ & $< 0.10$ & 1 & Strong lensing\\
$0.082 \pm 0.013$ & $< 0.10$ & 4 & X-ray \\
$0.50 \pm 0.09$ & $< 0.24$ & 2 & Weak lensing magnification \\
$0.48 \pm 0.16$ & $< 0.25$ & 3 & Weak lensing magnification \\
$0.20 \pm 0.03$ & $< 0.25$ & 4 & X-ray \\
\enddata
\tablerefs{
(1) \citet{tys95} ; (2) \citet{tay98}; (3) \citet{dye01};
(4) This study. 
}
\end{deluxetable*}

\subsection{Large scale configuration}

To illustrate the possible explanation that a merger might be 
taking place in this cluster and motivate why the X-ray derived 
mass should be lower than expected we employ a simple model. 
We assume that we have two perfectly spherical clusters aligned 
exactly along the line of sight having identical mass ($M_1$) and gas 
density ($n_1$). Since the thermal bremsstrahlung emissivity scales 
as $n^2$ we will measure a surface intensity $S(r) \propto n^2(r)$ and 
from this infer a gas density $n(r)$. In this example however we have 
$S(r) = 2 S_1(r) \propto 2 n_1^2(r)$ and hence we will conclude that  
$n(r) = \sqrt{2}~n_1(r)$. This means that we underestimate the total 
mass of the X-ray emitting gas by a factor $\sqrt{2}$.

From the hydrostatic equation (Eq. \ref{mass}) we can write 
$M(r) \propto \mathrm{d}(ln~n) / \mathrm{d}(ln~r) 
\propto n^{-1}~\mathrm{d}n / \mathrm{d}r$. From the gradient of the 
measured surface brightness 
$\mathrm{d}S / \mathrm{d}r \propto n(r)~\mathrm{d}n / \mathrm{d}r$ 
we infer a gradient on the gas density. In our scenario it is true 
that $\mathrm{d}S / \mathrm{d}r = 2~\mathrm{d}S_1 / \mathrm{d}r$ which 
yields $\mathrm{d}n / \mathrm{d}r = \sqrt{2}~\mathrm{d}n_1 / \mathrm{d}r$. 
For the total mass we will measure 
$M(r) \propto n^{-1}~\mathrm{d}n / \mathrm{d}r = 
n_1^{-1}~\mathrm{d}n_1 / \mathrm{d}r$. The actual mass of the 
cluster pair is of course $2~M_1$ which is twice what we measure. We 
have underestimated the total gravitating mass by a factor $2$. This 
is in agreement what we find in our comparison with gravitational 
lensing derived masses which would be able to measure the total mass 
accurately in this example.  

While this exact scenario is not very probable it shows that a 
close configuration of clusters will underestimate the X-ray mass, maybe 
by a factor as large as 2. In the above example the gas mass fraction 
would actually be overestimated which is the opposite of what we find. 
However if a merger was in fact taking place the hydrostatic equation 
would probably be quite inaccurate estimation of the mass. There 
would be other sources of pressure and probably the cluster would not 
be in equilibrium:  examples here might be magnetic fields, or
additional pressure support from non-thermal particles, likely to be 
accelerated in shocks that arise during a merger.  
Also since we do not detect two separate peaks in the surface intensity 
map, it is much more likely we have a lower density companion cluster or 
one merging irregular system.

\subsection{The dynamical state of Abell 1689}
There is a good evidence suggesting that Abell 1689 is undergoing 
a merger. We find that the X-ray measurement yields the value of mass 
that is low in 
comparison to gravitational lensing and the M-T relation for relaxed 
clusters, both by about a factor 2. We have shown that this may be 
the result of grouping or elongation along the line of sight.  
The cluster has a low gas mass fraction compared 
to other clusters, which is perhaps the result of large scale gas motion. 
The velocity dispersion of the member galaxies is very high and there 
are hints of subgroupings in the redshift space. 
The high number of spiral galaxies is unusual for a rich 
cluster and suggests a lower density region, where the spiral fraction 
would be higher, perhaps in front of or behind the main cluster.  
A state of complex dynamics is supported by the non-uniform temperature 
distribution and maybe most importantly by the variation in 
redshift of the X-ray emitting gas across the cluster.
Finally, recent measurements of the Sunyaev-Zeldovich 
effect by \citet{ben04} indicate that the inferred optical depth 
of the Comptonizing gas might actually be higher than one would 
infer from the simple spherically symmetric model obeying hydrostatic 
equilibrium, and a clumpy or elongated structure (aligned along the 
line of sight) would alleviate the discrepancy.  

Clearly, this cluster deserves further detailed studies;  
it is a very interesting potential target for the Astro-E2 
mission, where the X-ray calorimeter will provide an unprecedented 
spectral resolution, capable of clearly confirming the complex redshift 
structure of the intra-cluster gas, hinted by the XMM data.  If the cluster 
is indeed undergoing a merger, this might provide an environment where 
at least some fraction of the gas is accelerated (for instance, 
via shocks) to form a non-thermal distribution of particles, which might
produce radiation detectable in radio (via synchrotron emission), 
in hard X-rays (by Comptonizing the Cosmic Microwave Background), 
and in gamma rays, potentially detectable by the future mission GLAST.  

\section{Summary}
The superior effective area of the XMM-Newton telescope has allowed 
us to perform a detailed analysis of Abell 1689. Comparing with the data from 
ROSAT, Asca and Chandra we verified that the data are consistent with 
earlier observations. Importantly, there is no indication 
of any additional absorbing component in the XMM data, attributing the 
low-energy excess absorption found by \citet{xue02} in the Chandra data to 
uncorrected instrumental effects. The now available \verb1acisabs1 code 
successfully corrects for these effects. We confirm earlier findings that 
there is a large discrepancy, of a factor 2 or more, between mass estimates 
from gravitational lensing and X-ray derived mass using an unparameterized 
deprojection technique. Our analysis indicates that this discrepancy is true 
for the central part of the cluster, but also might be the case for 
the entire observable cluster;  our finding is in contrast to 
\citet{xue02}, who conclude that at large radii, there is no disagreement 
between the X-ray and lensing masses.  
Although the X-ray determined mass appears to be 
discrepant from the values determined from most lensing techniques, it 
seems to be in good agreement with that derived from 
weak gravitational shear.  We compare the gas mass of the cluster with the
total mass, and find that for Abell 1689, $f_{gas}=M_{gas}/M_{TOT}$ is
$0.072\pm 0.008$, significantly less than $0.108 \pm 0.014$, 
the value derived by \citet{all03} for 
10 dynamically relaxed clusters.  
Our calculation of the asymmetric temperature 
distribution of the cluster provides further evidence that this cluster 
is not in a relaxed state:  the lower than expected gas mass fraction 
is yet another piece of evidence.  We also present the first measurement 
of asymmetries in redshift for different regions of the ICM in the 
cluster determined from the X-ray data alone. A similar analysis was done  
by \citet{dup01} who claim a 3-$\sigma$ detection of bulk motion 
from Asca observations of the Centaurus cluster. 
We argue that the redshift variation detected here 
might be either due to line-of-sight clustering or 
possibly due to large bulk motions of the gas. Even though this 
cluster is clearly not as relaxed as might be expected from its 
apparent spherical form, we find the slope of the total mass in 
the central region to be in good agreement with what is expected 
from numerical simulations of structure formation. The density 
slope for $r < 90~ h^{-1}~ \mathrm{kpc}$ is $-1.3$. 
While our current understanding of the structure and dynamics of 
galaxy clusters is insufficient to put limits on the self-interaction 
of dark matter, better data and more accurate simulations appear 
promising for the future.  

\acknowledgments
We would like to thank Yasushi Ikebe and Steve Snowden for kindly 
helping in answering 
questions regarding XMM-Newton data reduction and calibration 
issues, and James Chiang for the suggestions regarding 
spectral deprojection. 
We are grateful to the XMM SAS software team for supplying the data 
reduction software. K. Andersson would also like to thank Per Carlson 
and Tsuneyoshi Kamae for making this research possible.  X-ray data used 
in this work was extracted from NASA's HEASARC archives.  We acknowledge the 
use of the Digitized Sky Survey, produced at the Space Telescope 
Science Institute under US Government grant NAG W-2166, based on the 
original photographic data obtained by the UK Schmidt telescope.  
This work was 
partially supported by the Alice \& Knut Wallenberg foundation via 
the Wallenberg Research Link, by NASA Chandra funds administered via 
SAO as GO1-2113X, and by Department of Energy contract DE-AC03-76SF00515 
to the Stanford Linear Accelerator Center.
We are grateful to the anonymous referee for many constructive 
suggestions and comments.

\clearpage


\clearpage


\end{document}